\documentclass[aps,twocolumn,showpacs]{revtex4}

\usepackage{amssymb}

\usepackage{amsmath}

\usepackage{epsf}

\begin{document}

\title{Condensation energy in Eliashberg theory -- from weak to strong coupling}

\author{Evelina Tsoncheva and  Andrey V. Chubukov}

\affiliation{
Department of Physics, University of Wisconsin, Madison, WI 53706}

\date{\today}

\begin{abstract}

We consider two issues related to the  
condensation energy in superconductors described by the 
Eliashberg theory for various 
forms of the pairing interaction, associated either with 
 phonon or electronic mechanisms of superconductivity. 
 First, we derive a leading 
correction to the BCS formula for the condensation energy to first order in the coupling $\lambda$.  Second, we
 show that at a given $\lambda$, the value of the condensation 
energy strongly depends on the functional form of the 
effective pairing interaction $\Gamma (\omega)$. 
\end{abstract}


\maketitle



A non-monotonic doping dependence of superconductivity in high $T_c$ cuprates revived the interest to the issue of the condensation energy in strongly coupled superconductors~\cite{hirsch,loram,scal_white,mike1,leggett,rob}. The condensation energy $E_c$ is the energy gain in a superconducting state compared to a normal state at the same $T$. In a BCS superconductor, $E_c$ smoothly increases below $T_c$ and at $T=0$ reaches $E_c^{BCS} = - V N_0 \Delta^2/2$, where $V$ is the volume, $\Delta$ is the superconducting gap and $N_0$ is the fermionic density of states \cite{bcs}. The decrease in the total energy upon pairing results from a fine competition between an increased kinetic energy and a decreased  

potential energy, both of which are much larger than $E_c$. Within BCS theory, the condensation energy is related to the jump of the specific heat at $T_c$ as  $C_s -C_n \approx 6.08 E_c/T_c$.

The BCS formula for the condensation energy is valid at weak coupling. 
Experimental discoveries of strong fermionic self-energy in the normal 
 state of the cuprates stimulated the studies of the condensation energy 
away from the BCS limit. In  earlier papers~\cite{rob}
 Haslinger and one of us analyzed the condensation energy  vs coupling 
 for spin-mediated $d-$wave superconductivity. It was found numerically 
 that the 
 condensation energy rapidly deviates from the BCS form as  coupling 
 increases, then saturates and then even decreases at strong coupling, 
 due to feedback on the bosonic spectrum from the pairing. The 
  crossover from weak to moderate coupling, however,
 has not been analyzed in detail in Ref\cite{rob}.

In the present paper, we consider
condensation energy in superconductors described by the 
Eliashberg theory, for various 
forms of the pairing interaction  $\Gamma (\omega) \propto (\omega^2_D + \omega^2)^{-\gamma/2}$ ($0<\gamma <2$), associated either with 
 phonon or electronic mechanisms of superconductivity (we discuss the 
examples below). 
 We address two issues that were not discussed in Ref. \cite{rob}.
First, we analyze, both analytically and numerically, the leading
correction to the BCS form of the condensation energy at weak coupling.
We show that the relative correction scales as a dimensional coupling $\lambda$
 and does not depend on the functional form of the pairing interaction (i.e., on $\gamma$).
Second, we compare the condensation energy for fixed $\lambda$, as a function
 of $\gamma$.  In the limit of  $\lambda \rightarrow \infty $, $E_c$ is the largest at $\gamma =2$,
 and for this $\gamma$ it actually diverges as $\log \lambda$.  
However, we found that this behavior only holds for extremely large $\lambda \sim 10^2-10^3$. At  moderate $\lambda$, we found numerically 
 an opposite trend:  a strong  increase of $E_c$ with decreasing $\gamma$.
 We understood this behavior analytically as originating already within the 
  BCS theory and coming from the strong $\gamma$ dependence of the prefactor 
 of the superconducting gap.

We consider a model in which low-energy fermions are 
interacting by exchanging either phonons or collective spin or charge 
fluctuations. This exchange gives rise to an attraction in one of 
 pairing channels, such that at $T=0$, the system is in the superconducting 
 state. 
At weak coupling, both phonon-mediated and collective mode-mediated pairings can be treated within BCS theory. At moderate and strong coupling, 
 we assume that the Eliashberg theory is valid~\cite{eliash_1}, i.e., that the self-energy is large, but it predominantly depends on frequency, and 
 vertex corrections can be  still neglected. 
Eliashberg theory can be straightforwardly justified for 
 phonon superconductors due to the smallness
 of the the sound velocity compared to the Fermi velocity~\cite{migdal,review}. 
For electronic mechanism of  pairing, it is justified on different reasons,
 but still, at low energies bosonic excitations 
 turn out to be slow modes   compared to electrons, i.e., the 
effective bosonic velocity is smaller than $v_F$.  We discuss the justification of the 
Eliashberg theory for electronic pairing mechanism in some detail 
at the end of the paper, after we discuss the results. 

The smallness of (real of effective) bosonic velocity compared to $v_F$ 
 implies that in the pairing problem, the momentum integration is factorized: the one over momenta transverse to the Fermi surface involves only fermions,
 while the one along the Fermi surface involves the bosonic propagator 
 $\chi (q_{\parallel}, q_{perp} =0, \omega)$. 
As a consequence, the theory operates with the effective 
``local'' pairing interaction $\Gamma_{\omega} \propto \int d q_{\parallel} 
\chi (q_{\parallel}, \omega)$. This interaction quite generally 
can be cast in the form~\cite{artem_unp} 
\begin{equation}
\Gamma_\omega = \left(1 + \omega^2/\omega^2_D\right)^{-\gamma/2}.
\label{new}
\end{equation}
where $\omega_D$ is characteristic frequency.  
The case $\gamma =2$ 
corresponds to phonon-mediated superconductivity, where
 $\omega_D$ is a Debye frequency. Other values of $\gamma$ correspond to electronic pairing. 
For instance, $\gamma =1/3$ corresponds to the pairing mediated by 2D 
 ferromagnetic Ising spin fluctuations or long-wavelength charge
 fluctuations~\cite{pepin}.
 In both cases, $\chi (q, \omega) \propto 
(\xi^{-2} + q^2 + |\omega|/q)$.
 Integrating over $q$, one immediately finds $\Gamma_{x} \rightarrow const$
 at $x \rightarrow 0$ and $\Gamma_x \propto (x/\xi^{-3})^{-1/3}$ at large $x$,
in agreement with (\ref{new}) for $\gamma =1/3$. The behavior at intermediate $x$ is more complex than in (\ref{new}) but this does not affect the physics. 
Similarly, $\gamma =1/2$ corresponds to the pairing mediated by 
 2D overdamped 
 spin or charge fluctuations peaked at a finite 
 momentum~\cite{italy,abanov_advances}, while 
$\gamma =1$ corresponds to the pairing mediated by propagating spin or 
 charge density waves. The case $\gamma \rightarrow 0$ describes all
 3D pairings mediated by either charge or spin fluctuations~\cite{theory}.  

The variables of the  Eliashberg theory are 
 the pairing gap $\Delta (\omega)$ and the 
quasiparticle renormalization factor $Z(\omega)$. They are related to 
 the fermionic self-energy 
$\Sigma (k, \omega)$ and the pairing vertex $\Phi (k, \omega)$ 
 by $\omega + \Sigma (\omega) = \omega Z (\omega), ~\Delta (\omega) = \Phi (\omega)/Z(\omega)$. The equation for $\Delta$ is  decoupled from that for $Z$ for 
 any form of $\Gamma_\omega$ and 
is obtained by straightforward extension of that for phonons~\cite{review}
\begin{equation}
\Delta_{\omega_{m'}} = \pi T \lambda \sum_{m=-\infty}^{\infty} \frac{\Delta_{\omega_m} - 
\Delta_{\omega_{m'}} \frac{\omega_m}{\omega_{m'}}}{\sqrt{\omega_m^2 + \Delta^2_{\omega_m}}} \Gamma_{\omega_{m'} - \omega_m} 
\label{3}
\end{equation}

The dimensionless $\lambda$ is defined as 
 $\lambda =({\bar g}/\omega_D)^\gamma$ where ${\bar g}$ is the effective 
 fermion-boson interaction. This definition implies that at $\omega_D 
\rightarrow 0$, the gap equation becomes independent 
on $\omega_D$ as it indeed should be
 as $\omega_D \rightarrow 0$ just implies that a pairing boson becomes gapless, 
 i.e., the pairing occurs near a quantum critical point (QCP).
  As in previous studies~\cite{review}, we fix ${\bar g}$ 
 to set the overall energy scale, and compute $E_c$ as a 
function of $\lambda$.  The choice to measure $E_c$ in units of 
 ${\bar g}$  is justified on 
the grounds that near QCP,  ${\bar g}$ does not critically depend on the 
distance to criticality.

Once $\Delta_{\omega_{m'}}$ is found by solving (\ref{3}), $Z_{\omega_{m'}}$ is immediately obtained from 

\begin{equation}
Z_{\omega_{m'}} = 1 +  \pi  T \lambda \sum_{m=-\infty}^\infty 
~\frac{1}{\sqrt{\omega^2_m + \Delta^2_{\omega_m}}}~\frac{\omega_m}{\omega_{m'}}
\Gamma_{\omega_{m'} - \omega_m}
\label{4}
\end{equation}

The expression for the condensation energy in the Eliashberg theory for phonons
 ($\gamma =2$) has been obtained by Wada and Bardeen and Stephen~\cite{bardeen_stephen}. The extension to arbitrary $\gamma$ is straightforward~\cite{artem_unp}.  In terms of  $D_{\omega_m} = \frac{\Delta_{\omega_m}}{w_m}$ and $\Gamma_{\omega_m}$
 the condensation energy reads
\begin{widetext}
\begin{eqnarray}
&&E_c = -N_0 \pi T \sum_{m} |\omega_m| 
\frac{(\sqrt{1 + D^2_{\omega_m}}-1)^2}{\sqrt{1 + D^2_{\omega_m}}} - 
N_0 \lambda \pi^2 T^2 \sum_{m} \sum_{m'} 
\frac{1 - sign (\omega_m \omega_m')}{\left(1 +((\omega_m - \omega_m')/\omega_D)^2
\right)^{\gamma/2}} \nonumber \\
&&
- \frac{N_0}{2}~\lambda \pi^2 T^2 \sum_{m} \sum_{m'} 
\left(\frac{(\sqrt{1 + D^2_{\omega_m}} - sign (\omega_m \omega_m')
{\sqrt{1 + D^2_{\omega_m'}}})^2}{\sqrt{1 + D^2_{\omega_m}} \sqrt{1 + D^2_{\omega_m'}}}\right)
~\frac{sign (\omega_m \omega_m')}{\left(1 +((\omega_m - \omega_m')/\omega_D)^2
\right)^{\gamma/2}} 
\label{5}
\end{eqnarray}
\end{widetext}
If we approximate $\Delta_\omega$ in the first term by a constant, 
this term yields a BCS condensation energy which at $T=0$ is 
$E_{c,BCS} = - N_0 \Delta^2/2$.
The combination of the second and the third terms in (\ref{5}),
 and the frequency dependence of $\Delta_\omega$, 
account for the corrections to BCS formula.
We analyzed both corrections and found that at small $\lambda$, the leading correction comes from the combination of second and third terms.
At $T=0$, the sum of these two terms reduces to 
\begin{widetext}
\begin{equation}
\delta E_c = -\frac{N_0 \lambda}{4} 
\int_0^\infty d \omega d \omega' 
\left(\frac{(\sqrt{1 + D^2_{\omega}} -
{\sqrt{1 + D^2_{\omega'}}})^2}{\sqrt{1 + D^2_{\omega}} \sqrt{1 + D^2_{\omega'}}}\right)~ \left(~\frac{1}{\left(1 +((\omega - \omega')/\omega_D)^2
\right)^{\gamma/2}} -~\frac{1}{\left(1 +((\omega + \omega')/\omega_D)^2
\right)^{\gamma/2}}\right)
\label{5_1}
\end{equation}
\end{widetext}
Simple considerations show that the dominant contribution to the 
 double integral comes from the range when either $\omega \sim \omega_D$,
 and $\omega' \sim \Delta$, or vice versa. Since at weak coupling, $\Delta << \omega_D$, one can expand in the smaller frequency in the last term in (\ref{5_1}).
One then obtains from (\ref{5_1}),
\begin{eqnarray}
&&\delta E_c = - N_0 \lambda \frac{\gamma}{\omega^2_D}~
\int_0^\infty  \frac{d \omega \omega}{\left(1 +(\omega/\omega_D)^2
\right)^{1+ \gamma/2}} \nonumber \\
&&~\int_0^\infty d \omega' \omega' \frac{(\sqrt{1 + D^2_{\omega'}} -1)^2}{\sqrt{1 + D^2_{\omega}}}
\label{5_2}
\end{eqnarray}
The last integral is the same one that yields a BCS condensation energy.
Evaluating the first integral, we obtain 
\begin{equation}
\delta E_c  =  \lambda E_{c,BCS} 
\label{5_3}
\end{equation}
such that 
\begin{equation}
E_c = -N_0 \frac{\Delta^2}{2} \left(1 + \lambda \right)
\label{6}
\end{equation}

Note that the relative correction to the BCS result depends only on $\lambda$, but not on $\gamma$. 

Eq. (\ref{6}) for $E_c$ can be formally obtained if one 
 assumes that the quasiparticle renormalization factor $Z (\omega)$ 
can be approximated by a constant, and this constant does not change between
 the superconducting and the normal state, 
where $Z = 1 + \lambda$~\cite{review}.
  For a constant $Z$, the full Green's function differs from the BCS result only by the renormalization $\epsilon_k \rightarrow \epsilon^*_k = \epsilon_k/Z$. 
Since the condensation energy is the integral over $\epsilon_k$ of the
 product of the Green's function and the self-energy, the effect of $Z$ can be
 absorbed into $d \epsilon_k = Z~ d \epsilon^*_k$. Obviously then 
$E_c = E_{c,BCS} \left(1 + \lambda \right)$ as in Eq. (\ref{6})~\cite{review}. 
 It is not a'priori clear, however, whether 
 the change in $Z$ between the normal and the superconducting state can be neglected, as, by  rough estimates, the contribution to $E_c$ from
 $Z_{sc} - Z_{n}$ is of the same order as in (\ref{5_3}).  We checked this explicitly for $\gamma =2$ using the Bardeen-Stephen formula for $E_c$  and found that the correction to $E_c$ due to  $Z_{sc} - Z_{n}$ is small, 
 because of an extra frequency integration involved,  and scales as 
$  \lambda E_{c,BCS}* (\Delta/\lambda \omega_D)^2 \ll \lambda E_{c,BCS}$. Our
 explicit calculation of $\delta E_c$ above shows that this smallness 
 survives for all $\gamma$.  

\begin{figure}[tbp]

\begin{center}

\epsfxsize=0.9 \columnwidth

\epsffile{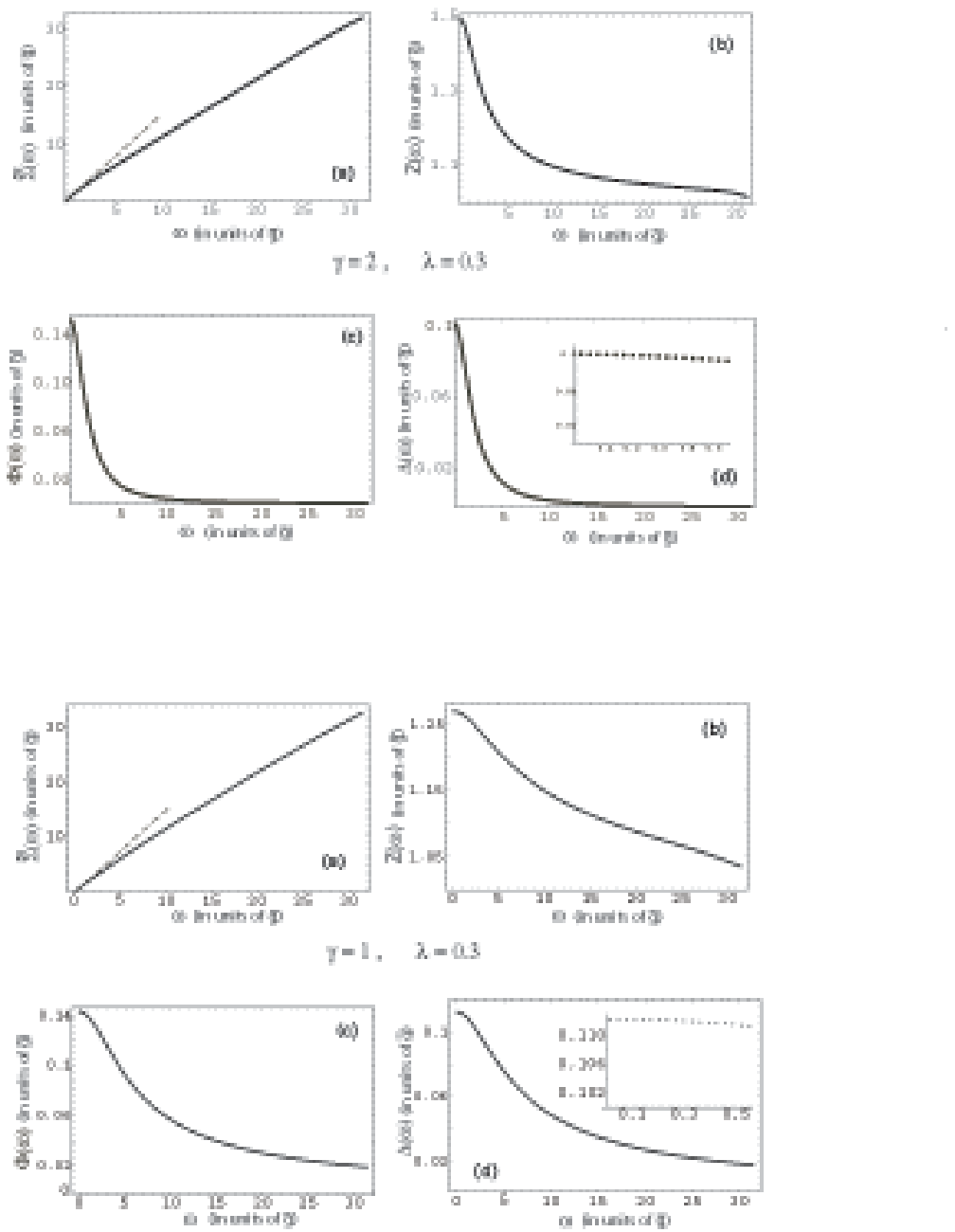}

\end{center}

\caption{The results for  (a) $~\Sigma_{\omega_m}$,  (b) $Z_{\omega_m}$,  (c) $\Phi_{\omega_m}$ and (d) $\Delta_{\omega_m}$. Upper panel, 
$\gamma =2$, lower panel $\gamma =1$. In both cases, we set
 $\lambda = 0.3$. The dashed lines in the plots for $~\Sigma_{\omega_m}$ indicates that $\Sigma_{\omega_m}$ is linear in 
frequency at small $\omega_m$.  The insert in the figure for $\Delta_{\omega_m}$  shows that the gap is nearly a constant for small $\omega_m$.}

\label{fig1}

\end{figure}

\begin{figure}[tbp]

\begin{center}

\epsfxsize=0.8 \columnwidth

\epsffile{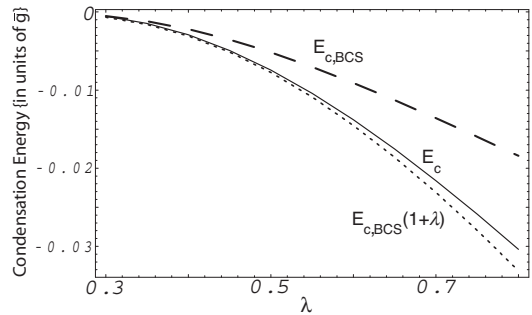}

\end{center}

\vspace{-0.75cm}

\caption{The condensation energy, $E_c$. Dashed, dotted and solid lines represent the BCS result, Eq. (6), and our numerical result for $E_c$ respectively. }

\label{fig2}

\end{figure}

To test this result, we computed $E_c$ numerically for a range of small
 to intermediate $\lambda$.

The equation for $\Delta_{\omega_m}$ was solved using an iterative computational method.   Two representative sets of results for 

$\Delta_{\omega_m}, Z_{\omega_m}, \tilde{\Sigma}_{\omega_m}$ and $\Phi_{\omega_m}$ for $\lambda = 0.3$ and $\gamma = 2$ and $\gamma = 1$ are shown in Fig.\ref{fig1}. 
  The condensation energy vs  $\lambda$ is plotted in Fig.\ref{fig2}. 
 At the lowest $\lambda <0.5$, we indeed reproduced the 
BCS result $E_{c,BCS} = -N_0 \frac{\Delta^2}{2}$. We see, however, that
 at intermediate $\lambda \leq 1$, the
 condensation energy deviates from the BCS formula, but 
closely  follows Eq. (\ref{6}).  At even larger $\lambda$,
 $E_c$ deviates even from the modified BCS expression.

\begin{figure}[tbp]

\begin{center}

\epsfxsize=0.8 \columnwidth

\epsffile{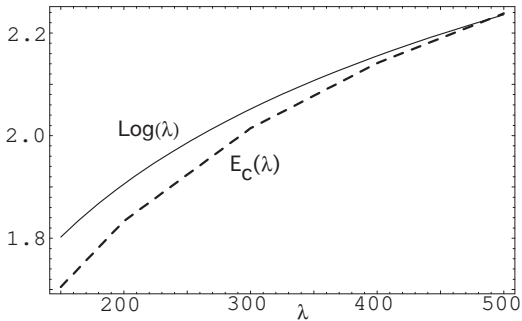}

\end{center}

\vspace{-0.75cm}

\caption{The  condensation energy $-E_c$ for $\gamma =2$ 
computed numerically for large $\lambda$ (solid line),
 and the analytical $E^{an}_c (\lambda) \propto \log \lambda$ within 
the same range (dashed line).}

\label{fig3}

\end{figure}

\begin{figure}[tbp]

\begin{center}

\epsfxsize= \columnwidth

\epsffile{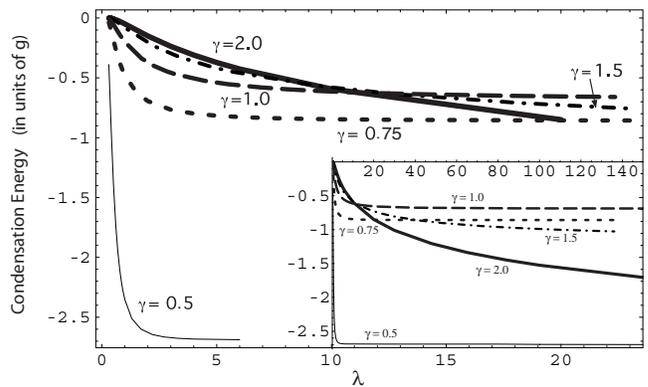}

\end{center}

\vspace{-0.75cm}

\caption{The condensation energy for various $\gamma$.
 The insert shows the large-$\lambda$ behavior of $E_c$. At weak
coupling, $|E_c|$ increases with decreasing $\gamma$, 
 while at very strong coupling it 
becomes the largest for $\gamma =2$. Observe that $E_c$ reaches saturation 
faster as $\gamma$ decreases.} 
\label{fig4}
\end{figure}

We next discuss the behavior of the condensation energy at large $\lambda$.
Here the dependence on $\gamma$ becomes crucial. First, we found from 
 numerical computation  that for $\gamma = 2$, 
 the condensation energy never saturates and keeps slowly increasing  with $\lambda$ (see Fig.\ref{fig3} and Fig.\ref{fig4}). 
 This can be easily understood analytically: using $\omega_D = {\bar g}/\sqrt{\lambda}$ and substituting $\lambda \Gamma_{\omega_m} = {\bar g}/(\omega^2_m + \lambda^{-1} {\bar g}^2)$ into Eq. (\ref{5}), one obtains that 
 the second term in (\ref{5}) diverges as $\log \lambda$~\cite{artem_unp}. 
 A more careful examination shows~\cite{artem_unp} 
that this divergence actually comes from the divergence of the 
 free energy of the normal state~\cite{comm1}
\begin{equation}
F_n = N_0 \pi^2 T^2 {\bar g}^2 \sum_{m,m'} \frac{1 - sgn (\omega_m) sgn (\omega_{m^{\prime}})}{(\omega_m - \omega_{m^{\prime}})^2 + 
\lambda^{-1}{\bar g}^2} 
\label{7}
\end{equation}

At $T=0$ and $\lambda \rightarrow \infty$ (i.e., $\omega_D \rightarrow 0$), the 2D integral in (\ref{7}) diverges as $\log \lambda$. 

At the same time, in the superconducting state, the free energy~\cite{artem_unp} 
\begin{eqnarray}
&&F_{sc} = 2 \pi T N_0 \sum_{\omega_m} |\omega_m| -
\frac{\omega_m^2}{\sqrt{\omega^2_m + \Delta^2_{\omega_m}}} - \pi^2 T^2 N_0
 \lambda \times \nonumber \\
&& \sum_{m,m'} \left(\frac{\omega_m \omega_{m'} + \Delta_{\omega_m} \Delta_{\omega_{m'}}}{\sqrt{\omega^2_m 
+ \Delta^2_{\omega_m}} \sqrt{\omega^2_{m'} + \Delta^2_{\omega_{m'}}}} -1 \right)~\Gamma_{\omega_{m^\prime}-\omega_m}
\label{8}
\end{eqnarray} 

is free from divergences as one can easily verify.  Obviously then,
$E_c= F_{sc} - F_n$ diverges as $\log \lambda$. 
In Fig.\ref{fig3} we plotted $E_c$ at large $\lambda$ vs $\log \lambda$.

We clearly see logarithmic behavior at  $\lambda \gg 1$. 

 The results for other $\gamma$ are presented  in Fig.\ref{fig4}.
For $\gamma <2$, the  condensation energy does not diverge at $\lambda = \infty$ as the frequency integral in $F_n$ in Eq. (\ref{7}) now converges. 
This implies that at large $\lambda$, $E_c$  decreases by magnitude 
when $\gamma$ decreases from $\gamma =2$. We indeed reproduced this behavior in our numerical calculations (see the insert in Fig.\ref{fig4}).
We, however, found numerically that $E_c (\gamma =2)$ becomes the largest by magnitude only at enormously large $\lambda \sim 10^3$. At moderate couplings, we actually found an opposite trend -- $|E_c|$ increases with decreasing $\gamma$,
 and the increase becomes quite strong when $\gamma$ becomes smaller than 1 (see the bulk of Fig. \ref{fig4}). This, e.g., 
implies that for same  $\lambda \geq 1$, in systems where the pairing is 
mediated by propagating optical phonons, the condensation energy in 3D  is larger than in 2D. It also implies that, 
 at the same moderate coupling $\lambda$, 
the  
2D pairing by ferromagnetic (Ising) spin fluctuations should 
yield larger $E_c$ than 2D pairing by Heisenberg antiferromagnetic spin fluctuations, and that for the  later, the condensation energy at a given $\lambda$ is larger for Ornstein-Zernike form of the spin susceptibility near $(\pi,\pi)$  than for flat spin susceptibility near $(\pi,\pi)$ 
(this flat susceptibility emerges in  RPA-based studies based on 
the measured  Fermi surface in $Bi2212$~\cite{mike}). 

\begin{figure}[tbp]
\begin{center}
\epsfxsize=0.8 \columnwidth

\epsffile{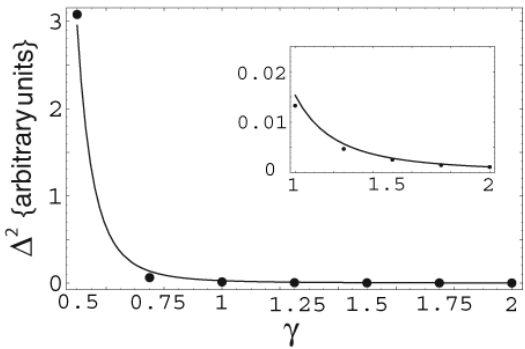}

\end{center}

\vspace{-0.75cm}

\caption{$\Delta^2 (\gamma)$ from Eq.(\protect\ref{11})
(line), together with the numerical results for ${\Delta_0}^2$ 
for  seven different $\gamma$ (dots). We used $\lambda = 0.3$. The inset shows a close view of $\Delta^2 (\gamma)$ for $1  \leq  \gamma  \leq  2$.}
\label{fig5}

\end{figure}

To understand the origin of this effect, we recall that for moderate couplings, one can use the modified BCS formula, Eq. (\ref{6}).
Since $\gamma$ enters this formula only through $\Delta$,  
large difference in $E_c$ between different $\gamma$ must be due to a difference in  the values of $\Delta$. To estimate  the $\gamma$ dependence of  
$\Delta$ at a given $\lambda \sim 1$, we just solve the 
gap equation to a logarithmic accuracy. Approximating $\Delta_\omega$
 by a constant $\Delta_0$, we obtain from Eq. (\ref{3})
\begin{equation}
\frac{1}{\lambda} = \int_{\bar{\Delta}_0}^{\infty} dx \frac{1}{x(x^2 +
 1)^\frac{\gamma}{2}}
\label{10}
\end{equation}

where ${\bar \Delta}_0 = \Delta_0/\omega_D$.
At $\gamma \rightarrow 0$ (i.e., when the pairing interaction tends to a constant), the integral diverges at the upper limit. This divergence is artificial 
as in reality the integration must not go beyond $x_{max} \sim E_F/\omega_D$.
We, however, will focus on finite $\gamma$, when the integral  converges.  
Carrying out the integration, we obtain an effective BCS formula
\begin{equation}
\Delta \sim  \omega_D~ e^{-\frac{1}{\lambda}}~ \exp\left[{\frac{\Psi(1) - \Psi(\frac{\gamma}{2})}{2}}\right]
\label{11}
\end{equation}
where $\Psi(x)$ is the digamma function.
 The first term in the r.h.s. of (\ref{11}) is the conventional weak-coupling result for a phonon superconductor. The dependence on $\gamma$ emerges 
through the second factor. 
We found that the dependence on $\gamma$ is 
 actually quite strong already at intermediate $\gamma$. 
 We plot our numerical $\Delta (\gamma)$ obtained for $\lambda = 0.5$ against Eq.\ref{11}) in Fig.\ref{fig5}.
 We see that the agreement between numerical and analytic $\Delta_0^2$ is quite good. As, to a very good accuracy,  the condensation energy for $\lambda =0.5$ 
 scales as $E_c \propto \Delta_0^2$, the good agreement in Fig. (\ref{fig5}) 
 implies that the origin of a very strong variation of $E_c$ with $\gamma$ 
 is indeed a strong $\gamma-$dependence of the magnitude of  $\Delta$.

Finally, we discuss the validity of the Eliashberg theory.
For phonon superconductors, Eliashberg theory is based on the smallness
 of the ratio of the electron mass and the ionic mass, or, alternatively, 
 with the smallness of the sound velocity $v_s$ compared to the Fermi velocity $v_F$~\cite{migdal}.
 The theory is valid as long as $\lambda v_s/v_F <1$,
 where $\lambda$ is the dimensionless coupling constant. On the scale of this parameter, one can neglect vertex corrections which are small in $\lambda v_s/v_F$.
Due to the smallness of $v_s/v_F$, the dimensionless $\lambda$ doesn't need to be small, and one can analyze both weak 
 coupling $\lambda \leq 1$ and strong coupling $1 < \lambda < v_F/v_s$ limits.

For pairing mediated by electronic collective modes, Eliashberg theory is justified on different grounds. The velocities of  collective modes are of the same order as the Fermi velocity, hence a direct argument based on $v_s/v_F$ ratio would not work. It turns out, however, that when the velocities of the two modes are comparable, the renormalization of the bosonic propagator due to Landau damping mechanism (which was unimportant for phonons) 
becomes relevant, and for $\lambda \geq 1$, collective modes
 become diffusive excitations at typical momenta and frequencies relevant to pairing. This  ensures that  at small frequencies, bosons become
 slow  modes compared to fermions. This is what Migdal theorem requires on physical grounds, as vertex corrections 
involve virtual processes in which fermions are forced to oscillate at 
 bosonic frequencies. Once bosons are slow mode compared to electrons, 
 fermions oscillate far from their own mass shell, and vertex corrections are reduced. The  calculations indeed
 show that at $\lambda \geq 1$, 
 vertex corrections are much smaller than $d\Sigma (\omega)/d\omega$
~\cite{abanov_advances,pepin}. For example, for the pairing by 
2D ferromagnetic  and incommensurate spin fluctuations, 
$d\Sigma (\omega)/d\omega = \lambda$ at large $\lambda$, while 
vertex corrections remain $O(1)$~\cite{pepin}. 
For the pairing by 2D antiferromagnetic spin fluctuations, 
vertex corrections grow with $\lambda$, but
 only as $\log {\lambda}$~\cite{abanov_advances}. In both cases, there is a wide range of $\lambda >1$ where frequency-dependent self-energy is already large, i.e., the system falls into the strong coupling regime, but vertex corrections still can be safely neglected. In this regime, Eliashberg theory is valid. Note, however, that unlike the case of electron-phonon interaction, Eliashberg theory for electron-electron interaction is {\it only} applicable at $\lambda \geq 1$. 

There are two other complications for collective mode-mediate pairing. First,
 the interaction is generally momentum-dependent, and the pairing is not of $s-type$. This does not invalidate Eliashberg theory as vertex corrections still remain small, and the momentum 
 integration in the gap equation is still factorized. As a result, $\Delta_{k, \omega} \approx \Delta_k \Delta_\omega$, and 
 the momentum dependence of the gap only affects the functional form of $\Gamma_\omega$, but does not change the structure of the equation for $\Delta_\omega$.
Second complication is that the propagator of a collective 
 mode by itself changes 
once the system enters a superconducting state. 
This, roughly, implies that $\gamma$ changes with $T/T_c$. 
In the analysis above we focused on $T=0$, assuming that all feedback effects are already incorporated into $\Gamma_\omega$.

To conclude, in this paper we computed the condensation energy $E_c$ for Eliashberg superconductors for a wide range of coupling strengths and 
for several different physically motivated forms of the pairing interaction
 (parameterized by $\gamma$), representing both phonon $(\gamma =2)$
 and electronic collective mode  mediated superconductivity $(0<\gamma <2)$. 
 We found  analytically and confirmed and numerically
 that for weak coupling,  
$E_c = E_{c,BCS} (1 + \lambda)$, where 
 $E_{c,BCS} = - N_0 \Delta^2/2$  is BCS result. This result holds for all $\gamma$. 
We also found that in the opposite limit of very large $\lambda$, 
$E_c$ diverges as $\log \lambda$ if the 
pairing is mediated by phonons, but remains finite for electronic mechanisms of pairing and decreases with decreasing $\gamma$. 
However, for moderate $\lambda \geq 1$, we found an opposite trend --
 the condensation energy at a given $\lambda$ strongly increases with decreasing $\gamma$.  We argued that this behavior holds in the regime where 
$E_c \propto \Delta^2$, and  is a consequence
 of the strong $\gamma$ dependence of the pairing gap $\Delta$.
 We computed $E_c (\gamma)$ analytically and numerically and  obtained a very good agreement between the two  results.

We acknowledge useful discussions with A. Abanov, B. Altshuler and E. Yuzbashyan. The research was supported by NSF DMR 0240238.


\begin{thebibliography}{9}
\bibitem{comm1} Eqs. (\ref{7}) and (\ref{8}) require clarifications. Both normal and superconducting Free energies obtained from 

 Luttinger-Ward functional contain an extra term $F_{ex} =  - N_0 \left[2 \pi T \sum_{m} |\omega_m| + 
 \lambda \pi^2 T^2 {\bar g}^2 \sum_{m,m'} \Gamma_{\omega_m - \omega^\prime_m}\right]$. The first term in this extra piece  does not depend on $\lambda$ and formally diverges at any $\lambda$ and any $T$, the second diverges as power of $\lambda$ at $\lambda \rightarrow \infty$ due to self-action piece with $m=m^\prime$. This $F_{ex}$ is, however, 
 irrelevant from physics perspective as the first term is  independent on the interaction, while the second term is just the frequency integral of the interaction potential, and it does not distinguish between superconducting and normal states.

 We therefore can safely subtract $F_{ex}$ from the Free energy. This leads to Eqs. (\ref{7}) and (\ref{8}).
\end{thebibliography}
\end{document}